\newif\ifemulate
\newif\ifastroph

\emulatetrue

\astrophtrue

\ifemulate
	\documentclass[iop,numberedappendix]{emulateapj}
\else
	\documentclass[manuscript]{aastex}
\fi

\newcommand{\myemail}{mulders@lpl.arizona.edu}

\slugcomment{Accepted by ApJ}

\shorttitle{Viscous snow line and water delivery}
\shortauthors{Mulders et al.}

\usepackage{natbib,graphicx,xspace,amsmath}
\newcommand{\SL}{\ensuremath{R_{\rm SL}}\xspace}
\newcommand{\Myr}{\ensuremath{\rm Myr}\xspace}
\newcommand{\Msunyr}{\ensuremath{M_\odot / {\rm yr}}\xspace}
\newcommand{\Mm}{\ensuremath{\dot{M}_m}\xspace}

\ifemulate
	\newcommand{\figa}{left\xspace}
	\newcommand{\figb}{right\xspace}
	\newcommand{\figwidth}{0.49}
\else
	\newcommand{\figa}{top\xspace}
	\newcommand{\figb}{bottom\xspace}
	\newcommand{\figwidth}{0.7}
\fi

\begin{document}

\title{
The snow line in viscous disks around low-mass stars: implications for water delivery to terrestrial planets in the habitable zone
}

\author{Gijs D. Mulders}
\affil{Lunar and Planetary Laboratory, The University of Arizona, Tucson, AZ 85721, USA}
\email{\myemail}

\author{Fred J. Ciesla}
\affil{Department of the Geophysical Sciences, The University of Chicago, 5734 South Ellis Avenue, Chicago, IL 60637}
\author{Michiel Min}
\affil{Astronomical Institute “Anton Pannekoek”, University of Amsterdam, The Netherlands}
\author{Ilaria Pascucci}
\affil{Lunar and Planetary Laboratory, The University of Arizona, Tucson, AZ 85721, USA}

\begin{abstract}
The water ice or snow line is one of the key properties of protoplanetary disks that determines the water content of terrestrial planets in the habitable zone. 
Its location is determined by the properties of the star, the mass accretion rate through the disk, and the size distribution of dust suspended in the disk.
We calculate the snow line location from recent observations of mass accretion rates and as a function of stellar mass. By taking the observed dispersion in mass accretion rates as a measure of the dispersion in initial disk mass, we find that stars of a given mass will exhibit a range of snow line locations. At a given age and stellar mass, the observed dispersion in mass accretion rates of 0.4 dex naturally leads to a dispersion in snow line locations of $\sim$0.2 dex. 
For ISM-like dust sizes, the one-sigma snow line location among solar mass stars of the same age ranges from $\sim$2 to $\sim$5 au. For more realistic dust opacities that include larger grains, the snow line is located up to two times closer to the star.
We use these locations and the outcome of N-body simulations to predict the amount of water delivered to terrestrial planets that formed in situ in the habitable zone.
We find that the dispersion in snow line locations leads to a large range in water content.
For ISM-like dust sizes, a significant fraction of habitable-zone terrestrial planets around sun-like stars remain dry, and no water is delivered to the habitable zones of low-mass M stars ($<0.5 M_\odot$) as in previous works.
The closer-in snow line in disks with larger grains enables water delivery to the habitable zone for a significant fraction of M stars and all FGK stars. 
Considering their larger numbers and higher planet occurrence, M stars may host most of the water-rich terrestrial planets in the galaxy if these planets are able to hold on to their water in their subsequent evolution.

\end{abstract}


\section{Introduction}
The leading explanation for the delivery of water to Earth is that water-bearing, asteroid-like bodies from beyond the snow line were gravitationally scattered inward and accreted by the planet during its growth \citep[e.g.][]{Morbidelli:2000ex}. 
If the same mechanism also operates in extra-solar planetary systems, the water content of potentially habitable terrestrial planets that formed in situ depends on both the snow line location and the extent to which water-bearing materials were scattered inwards. For lower-mass M dwarf stars, N-body simulations indicated that terrestrial planets in the habitable zone would be dry \citep{2007ApJ...669..606R, 2007ApJ...660L.149L}, in contrast to planets that formed farther out in the disk and migrated to their current locations \citep{Ogihara:2009cu}. In a previous paper \citep{Ciesla:2015ha}, we explored how the inclusion of planetesimals and more comet-like bodies in these simulations can enhance volatile delivery around these low-mass stars. In that work, the location of the snow line was a critical factor in determining how much water would be delivered to forming planets.  Here, we revisit the location of the snow line based on recent protoplanetary disk observations, and explore its impact on water delivery to habitable zone terrestrial planets.

\begin{figure*}
	\includegraphics[width=\figwidth\linewidth]{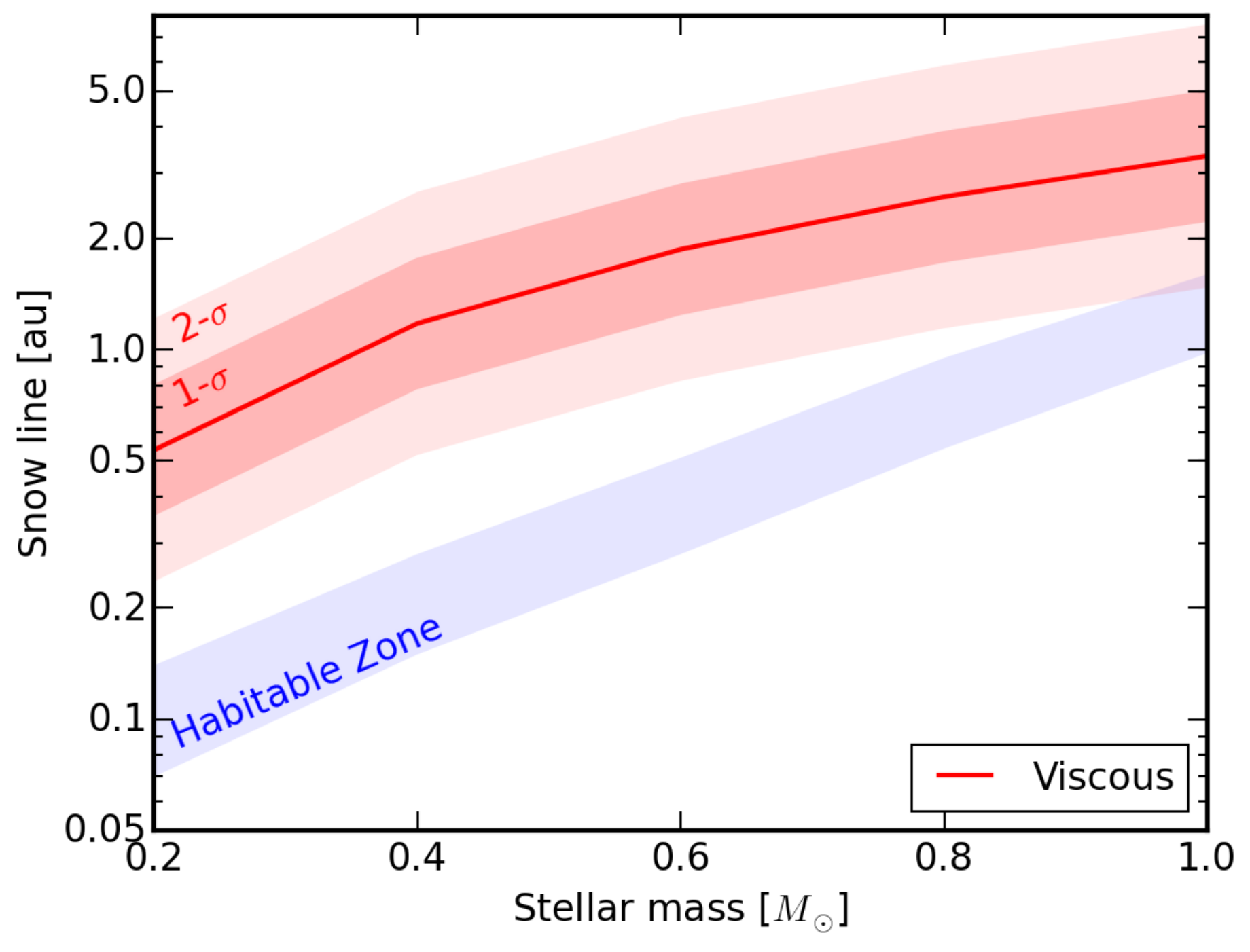}
	\includegraphics[width=\figwidth\linewidth]{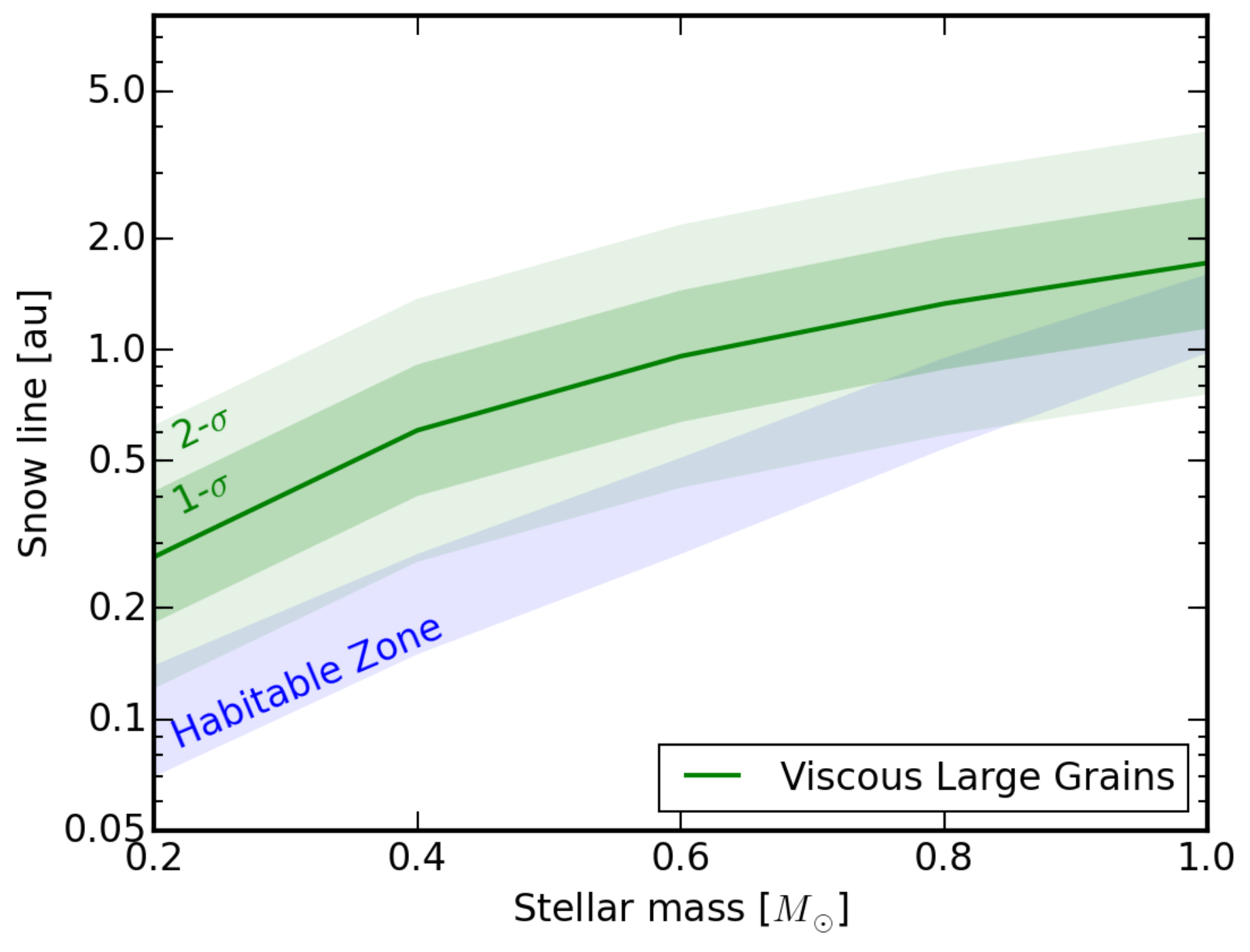}	

	\caption{Location of the snow line for a range of stellar masses for a disk of small ISM-like grains (\figa panel) and larger grains more representative of protoplanetary disks (\figb). The green/red line, dark-shaded area and light-shaded area show the median, $\pm 1\sigma$, and $\pm 2 \sigma$ location of the snow line, respectively, calculated from the observed distribution of mass accretion rates using Eq. \ref{eq:analytical}.
	\label{fig:diag}
	}
\end{figure*}

In the Solar System, the location of the snow line at the time of planetesimal formation has been inferred from the transition between hydrous and anhydrous asteroids to be $\sim 2.5$ au \citep{Abe:2000vr}. There is, however, a considerable uncertainty in this location if asteroids have been scattered into their current locations \citep[e.g.][]{2012M&PS...47.1941W,DeMeo:2014hk}.
There are no direct measurements of the location of a water snow line outside our solar system, hence we do not know if it is typical for sun-like stars. The Solar System snow line location of 2.5 au has been taken as a reference point in previous studies, and used to estimate the location around other stars of various masses \citep[e.g.][]{2007ApJ...669..606R}. The exact location of the snow line will be determined by the combined irradiative and viscous heating in the disk \citep[e.g.][]{Davis:2005ho}.  If mass accretion occurs through the vertical extent of the disk, the mid plane temperatures at the relevant locations are set by the release of gravitational potential energy for mass accretion rates $>10^{-10} \Msunyr$, and the location of the snow line is a strong function of mass accretion rate \citep{2007ApJ...654..606G,Oka:2011jh}. 

The dependence of mass accretion rate on stellar mass has been well-established, following a roughly quadratic relationship \citep[e.g.][]{Muzerolle:2003ha,2004AJ....128.1294C,2005ApJ...625..906M, 2006A&A...452..245N}, but these relations have not been used to calculate the location of the snow line around low-mass stars. On top of that, there is an intrinsic dispersion in mass accretion rates of $\sim 0.4$ dex around this relation \citep{2014A&A...561A...2A}. If this intrinsic dispersions reflects a range in initial disk masses, which is of the same order, 0.5 dex \citep{Armitage:2003dc}, we expect an associated dispersion in snow line locations (Figure \ref{fig:diag}), leading to enhanced or reduced water delivery among stars of similar mass. 

In addition to the mass accretion rate, the location of the snow line in a protoplanetary disk around a sun-like star is very sensitive to the dust opacity \citep{2011Icar..212..416M, Oka:2011jh}.  While previous work on the snow line location in the solar nebula had mainly used small sub-micron sized grains \citep[e.g.][]{Davis:2005ho,2007ApJ...654..606G}, there is abundant observational and theoretical evidence that grains in protoplanetary disks quickly grow to millimeter and centimeter sizes \citep[e.g.][]{2014prpl.conf..339T}, depleting the amount of small grain by orders of magnitude \citep[e.g.][]{2006ApJS..165..568F}. Lower dust opacities trap viscous heat less efficiently, resulting in a cooler disk mid plane with a snow line closer to the star, enhancing water delivery.

In this work, we explore the range of snow line locations that may be expected for stars of different masses, factoring in variations in disk mass accretion rate and dust opacities that have been inferred for real disks (\S \ref{sec:SL}) and how
this would impact the water content of terrestrial planets in the Habitable zone (\S \ref{sec:nbody}).

\begin{table*}
	\centering

	\begin{tabular}{l l l l l l}\hline\hline
	$M_*$ 			& $\log \dot{M}$  		& \multicolumn{2}{c}{\SL (1-$\sigma$)} & \multicolumn{2}{c}{\SL} \\
	 				& 						& small grains 	& large grains & MS & pre-MS \\
	{}[$M_\odot $]	& [$\log M_\odot /{\rm yr}$]		& [au] & [au] & [au] & [au] \\
	\hline
	$1.0$ & $-7.5 \pm 0.4$ & $3.3$ ($2.2$...$5.0$) & $1.6$ ($1.1$...$2.4$) & 2.5 & 1.35 \\
	$0.8$ & $-7.7 \pm 0.4$ & $2.6$ ($1.7$...$3.9$) & $1.2$ ($0.82$...$1.9$) & 1.23 & 1.15 \\
	$0.6$ & $-7.9 \pm 0.4$ & $1.9$ ($1.2$...$2.8$) & $0.9$ ($0.59$...$1.3$) & 0.61 & 0.95 \\
	$0.4$ & $-8.3 \pm 0.4$ & $1.2$ ($0.78$...$1.8$) & $0.6$ ($0.37$...$0.84$) & 0.32 & 0.8 \\
	$0.2$ & $-8.8 \pm 0.4$ & $0.5$ ($0.36$...$0.81$) & $0.2$ ($0.17$...$0.38$) & 0.16 & 0.5 \\
		
	\hline\hline\end{tabular}
	\caption{
	Mass accretion rates and location of the viscous snow line as a function of stellar mass used in this work, including one-sigma ranges. For comparison, the last two column shows the snow line locations calculated from the stellar luminosity used in previous work: MS (main-sequence, \citealt{2007ApJ...669..606R}) and pre-MS \citep{Ciesla:2015ha}.
	}
	\label{tab:SL}
\end{table*} 

\section{Snow line location in a viscous disk}\label{sec:SL}

The location of the snow line, \SL, in a steady-state viscous protoplanetary disk, assuming it is optically thick to its own radiation, is given by \cite{2011Icar..212..416M}, their Eq. (11):
\begin{equation}
	\label{eq:analytical}
	\SL =  \left(\frac{3 \mu m_p (GM_\star)^{3/2}\dot{M}^2\kappa_R}{128 \pi^2 k_b \sigma_{\rm SB} T_{\rm ice} f \alpha}\right)^{2/9},
\end{equation}
where $\mu$ is the mean molecular weight, $m_p$ the proton mass, $G$ the gravitational constant, $\dot{M}$ the mass accretion rate, $M_\star$ the stellar mass, $\kappa_R$ the Rosseland mean opacity, $k_b$ the Boltzmann constant, $\sigma_{\rm SB}$ the Stefan-Boltzmann constant, $T_{\rm ice}$ the temperature where water ice condenses, $f$ the gas-to-dust ratio, and $\alpha$ the turbulent mixing strength. This equation is based on the estimate of the mid plane temperature of a viscous disk by \cite{Hubeny:1990kr}. \cite{2011Icar..212..416M} showed that this equation is in good agreement with the snow line computed through detailed radiative transfer modeling for a sun-like star. Even though the role of irradiation becomes larger when using the pre-main sequence luminosity and when considering lower mass disks around lower mass stars, we have verified in Appendix \ref{app:mcmax} that viscous heating indeed dominates the thermal budget of the disk mid plane for the range of stellar masses and disk mass accretion rates explored in this paper.

The ice sublimation temperature is thought to lie between 150 and 170 K \citep[e.g.][]{Podolak:2004er}, depending on the water vapor pressure in the disk. The radiative transfer models in \cite{2011Icar..212..416M} include a detailed treatment of solid sublimation as described in \citep{2009A&A...506.1199K}. \cite{2011Icar..212..416M} show that, for a large range of mass accretion rates ---  and hence surface densities and partial vapor pressures --- the location of the snow line is well predicted by assuming a single sublimation temperature of $T_{\rm ice}=160 K$. We therefore do not solve the radial diffusion equation \citep[e.g.][]{2006Icar..181..178C}, as we expect the variations of the partial water pressure by inward drift of evaporating ices to be within the tested range and not lead to large variations in the location of the snow line.

Grouping all constants together, taking $\mu=2.3$, we obtain the more manageable expression
\begin{equation}\label{eq:SLau}
\begin{split}	
	\SL= 2.1 ~{\rm au}~ & \left(\frac{M_\star}{M_\odot}\right)^{1/3}  
	\left(\frac{\dot{M}}{10^{-8} \Msunyr}\right)^{4/9}  \\
	& \left(\frac{\kappa_R}{770 ~{\rm cm}^2/g}\right)^{2/9} 
	\left(\frac{f}{100}\right)^{-2/9} \\
	&\left(\frac{\alpha}{0.01}\right)^{-2/9}  
	\left(\frac{T_{\rm ice}}{160 ~K}\right)^{-10/9}.
\end{split}
\end{equation}
In the remainder of the section we explain the origin and applicability of the parameters in equation \ref{eq:SLau} that lead to the range of snow lines shown in Figure \ref{fig:diag}.

\subsection{Disk mass accretion rate}
The mass accretion rate scales roughly with stellar mass squared \citep[e.g.][]{Muzerolle:2003ha,2004AJ....128.1294C,2005ApJ...625..906M, 2006A&A...452..245N}, with an observed dispersion of up to an order of magnitude around this relation. We will use the observed mass accretion rates from \cite{2014A&A...561A...2A} since these have the smallest observed dispersion, and we will explain in section \ref{sec:spread} how we treat this dispersion.

The \textit{median} mass accretion rate for the $\sim 3 \pm 1$ Myr old \citep{2008hsf2.book..295C} Lupus star forming region, over a range of stellar masses from $0.03~M_\odot$ to $1.0 M_\odot$, is given by \cite{2014A&A...561A...2A}:
\begin{equation}
	\label{eq:alcala}
	\dot{M}_{\rm Lupus}(M_\star) = 5.6 \cdot 10^{-9} \left[\frac{M_\star}{M_\odot}\right]^{1.81} ~\Msunyr.
\end{equation}

We derive a generic expression for the median mass accretion rate as a function of time by appending this equation with a factor $t^{-3/2}$ to take into account the age of the cluster:
\begin{equation}
	\label{eq:Mdotmedian}
	\Mm(M_\star,t) = 3 \cdot 10^{-8} \left[\frac{M_\star}{M_\odot}\right]^{1.81} \left[\frac{t}{\Myr}\right]^{-3/2} ~\Msunyr.
\end{equation}
This accretion rate is consistent with the derived accretion rate for the 1 Myr old Taurus region of $\sim 10^{-8} \Msunyr$ for $\sim 0.5 ~M_\odot$ stars \citep{1998ApJ...495..385H}, and matches well with the observed decay of the mass accretion rate between 0.3 and 30 Myr \citep[][their Figure 2]{SiciliaAguilar:2010jn}. The exponent of $\eta=1.5$ comes from a viscously evolving disk with fixed temperature \citep{2006ApJ...648..484H}, and matches well with the analytical model of \cite{2009ApJ...705.1206C} that includes a changing disk temperature ($\eta=20/13=1.54$) which is more similar to the steady state accretion disk models employed in deriving Eq. \ref{eq:analytical} by \cite{2011Icar..212..416M}. Eq. \ref{eq:Mdotmedian} is only valid at time scales much longer than the viscous time scale, and starts significantly over-predicting the mass accretion rate at $t<0.3$ Myr for $\alpha=0.01$ compared to the \cite{2009ApJ...705.1206C} disk model.

\subsection{Range of snow line locations}\label{sec:spread}
A dispersion in disk mass accretion rate at a given age will lead to a dispersion in snow line locations. Observed mass accretion rates show a large dispersion at any given stellar mass. Part of this dispersion may be attributed to variations in the accretion flow onto the star \citep{2012MNRAS.427.1344C}, and use of secondary tracers of accretion such as emission lines \citep{2012A&A...548A..56R}. An intrinsic dispersion may reflect an age spread and variations in the initial disk mass \citep[e.g.][]{1998ApJ...495..385H}. The tightest constraint on the intrinsic dispersion currently comes from \cite{2014A&A...561A...2A}. Using the X-shooter spectrograph, the authors compute the mass accretion rates using the UV excess emission, a direct tracer of accretion, in the Lupus star forming region over a range of stellar masses from $0.03~M_\odot$ to $1.0 M_\odot$. The dispersion in mass accretion rates \textit{at a given stellar mass} is characterized by a log-normal distribution with a standard deviation $\sigma_{\dot{M}} = 0.4$ dex, a factor of two lower than in previous works. We assume this dispersion reflects a range of initial disk masses as proposed by \cite{1998ApJ...495..385H}.

This assumption is supported by two results. First, the diversity in protoplanetary disk properties, in particular the outer radius-disk mass relation \citep{Andrews:2010eg} and the exponential decay of the disk fraction versus time \citep[e.g.][]{Mamajek:2009de} argue for an intrinsic spread in initial disk masses. Since the inferred spread in disk masses from the time-dependent disks fraction is of the same order, $\sim 0.5$ dex \citep{Armitage:2003dc}, we take the observed dispersion in mass accretion rate to be the intrinsic one. Second, although accretion rates measured from spectral lines are known to be variable on time scales of days to years, this variation is smaller than the observed dispersion and driven by the rotational modulation of the accretion flow onto the star \citep{2014MNRAS.440.3444C}, and does not reflect an intrinsic variation in the \textit{disk} mass accretion rate. It is possible that disks show larger variations at time scales of decades or longer but smaller in magnitude than FUor outbursts, or that it reflect a large age spread in the Lupus cloud, in which case the intrinsic scatter could be smaller. 

With the assumption that the observed spread in mass accretion rates is the intrinsic one, the probability distribution of the snow line $R$ around the median, \SL (Eq. \ref{eq:SLau}), is given by a log-normal distribution:
\begin{equation}
	\label{eq:lognormal}
	f_{\rm SL}(R,\SL)= \frac{1}{\sqrt{2 \pi} ~R ~\sigma_R } \exp\left(-\frac{(\log R - \log \SL)^2}{2 \sigma_R^2}   \right)
\end{equation}
with a standard deviation $\sigma_R= 4/9* \sigma_{\dot{M}} \approx 0.18$ dex. This distribution is shown as the shaded regions in Fig. \ref{fig:diag}.

\subsection{Dust opacity}
Since a major source of uncertainty in the calculated location of the snow line is the dust opacity \citep{2011Icar..212..416M}, we explore two limiting cases: the first is where the grains are ISM-like in size as in most previous work, with a Rosseland mean opacity at temperatures relevant for water ice condensation of $\kappa_R = 570 ~{\rm cm}^2/g$, as in \cite{2011Icar..212..416M}; the second is where dust growth is assumed to occur, resulting in a size distribution from micron to centimeter-sized grains, yielding an opacity of $\kappa_R = 20 ~{\rm cm}^2/g$. Such a lower opacity will also increase the amount of solids available for planetesimal formation interior to the snow line, see \cite{2011Icar..212..416M} for details. Because the dependence of the snow line location on this opacity is relatively weak ($\SL \propto \kappa_R^{2/9}$, Eq \ref{eq:SLau}), the snow line moves in by about a factor of two between the two cases considered here (\figa and \figb panels of Fig. \ref{fig:diag}). Table \ref{tab:SL} lists the mass accretion and corresponding snow line locations for the range of stellar masses for the two opacities considered here. Figure \ref{fig:diag} shows the median snow lines for the same stellar mass range, compared to the recent habitable zone estimate from \cite{Kopparapu:2013fu}. Throughout this paper, we use a gas-to-dust ratio $f=66$ based on a condensation sequence at solar metallicity, following \cite{2011Icar..212..416M}.

\subsection{Snow line location}
For a solar mass star with small grains, the one-sigma range of snow line locations encompasses that of the Solar System, from $\sim$2 to $\sim$5 au. Its median location is consistent with the \cite{2009ApJ...705.1206C} disk model for a disk of the same mass accretion rate, age, and opacity. Due to the relatively weak scaling with mass accretion rate ($\dot{M}^{4/9}$), the two-sigma range of snow line location changes by a factor five, even if $\dot{M}$ changes by a factor of $40$. 
For the large grain case, the one-sigma locations range from $1.1$ to $2.4$ au, roughly encompassing the habitable zone. In the context of these models, if the solar system snow line was located at 2.5 au at 1 Myr, it is a one-sigma out-lier, and the majority of exoplanetary systems may form with a snow line closer to the star.

Around lower mass stars, the snow line moves closer to the star as $\SL \propto M_\star^{1.14}$ (Eqs \ref{eq:SLau}, \ref{eq:Mdotmedian}). The snow line around a $0.6 M_\odot$ star is closer in compared to that inferred by \cite{2008ApJ...673..502K}, who use a linear scaling of the mass accretion rate with stellar mass which is no longer supported by observations, leading to a hotter disk with a more distant snow line. Compared to the habitable zone, that scales with stellar mass roughly as $R_{\rm HZ} \propto \sqrt{L_{\star, \rm MS}} \propto M_\star^{1.75...2.0}$, the relative gap between snow line and habitable zone widens \citep[See also ][]{2007ApJ...660L.149L}. For the small grain case, the gap between the median snow line and the habitable zone is about a factor of five. For the large grain case, this is only a factor of two, and the two-sigma tail of the snow line distribution overlaps with the habitable zone.

\begin{figure}
	\includegraphics[width=\linewidth]{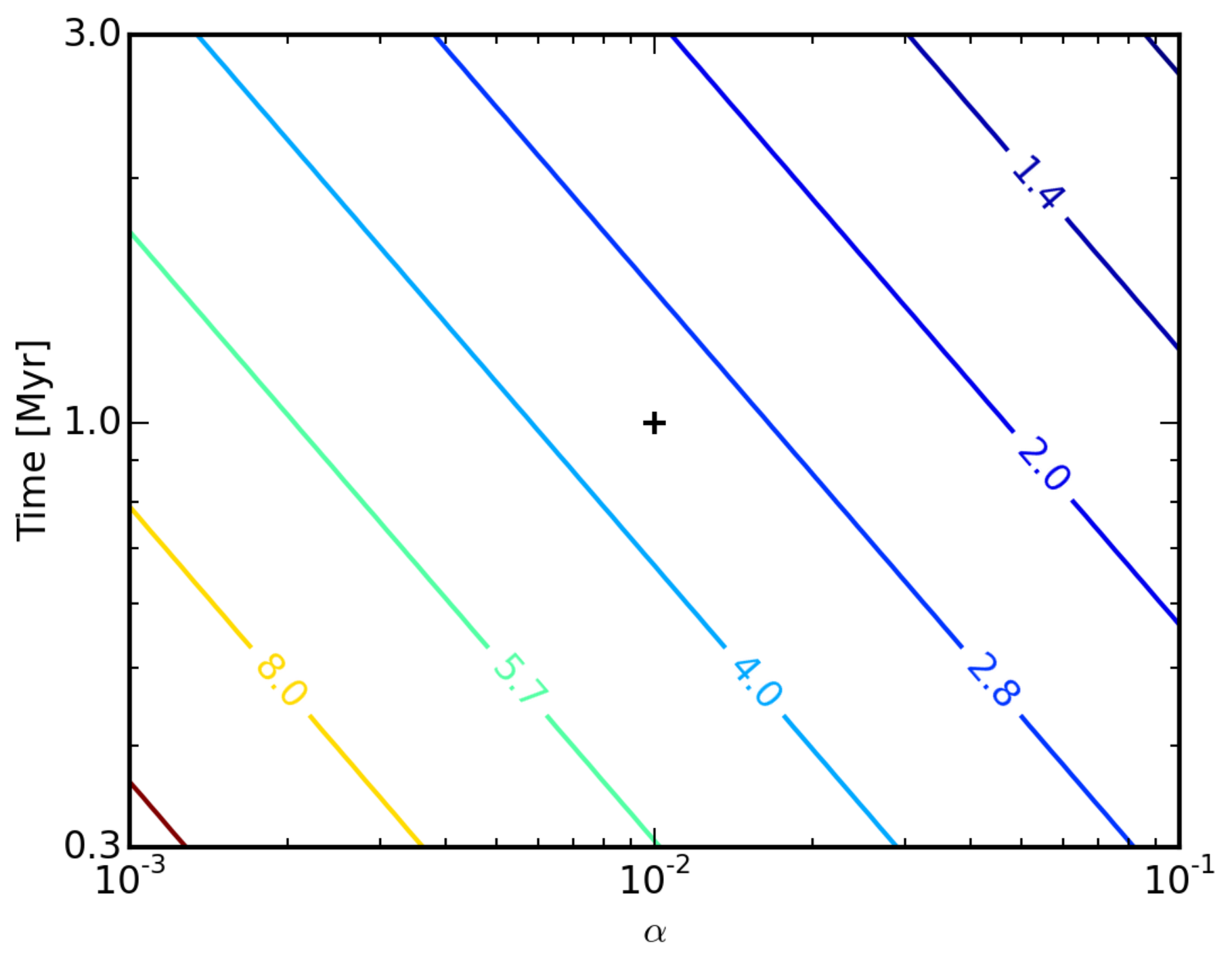}
	\caption{The location of the snow line in au around a solar mass star, for a range of the turbulent mixing strength $\alpha$ and the time $t$ at which the snow line is ``frozen in'' the planetesimal population. Our choice of $\alpha=0.01$ and $t=1$ Myr is shown by the central cross. In general, lower turbulent mixing strength and earlier times move the snow line outward. Towards higher turbulent mixing strength and later times it moves inward.
	\label{fig:timealpha}
	}
\end{figure}

\subsection{Disk viscosity and time of planetesimal formation}
Throughout this paper, we assume $t=1~\Myr$ and $\alpha=0.01$ for setting the location of the snow line. Both quantities are not well constrained observationally. The strength of viscosity in protoplanetary disks is inferred to be of order $\alpha \sim 0.01$ based on disk masses, sizes, and lifetimes \citep{1998ApJ...495..385H}. The time that defines the location of the snow line depends on both the planet formation time scale and the extent to which (icy) grains continue to be incorporated into larger bodies as the snow line moves in. The theoretical planetesimal formation time scale at $\sim 1 ~AU$ is of order $0.01-0.1 Myr$ \citep[e.g.][]{Wetherill:1993fp}, and the earliest differentiated bodies formed within $< 1 Myr$ \citep{Kleine:2009jv}, suggesting an early epoch of planetesimal formation. On the other hand, the age of the chondrules in chondritic parent bodies indicate planetesimal formation continued to take place over a 3 Myr time scale \citep[e.g.][]{2014prpl.conf..547J}. If water vapor remains present in the disk, an inward moving snow line may give rise to rapid, but late icy planetesimal formation \citep{2006ApJ...650L.139K}. The age of $t=1~\Myr$ is a compromise between two extremes. 

To give the reader an idea of how these uncertainties impact the location of the snow line, Figure \ref{fig:timealpha} shows how the location of the snow line varies around a sun-like star with a median mass accretion rate following equation \ref{eq:SLau}. Earlier times and lower turbulent mixing strength correspond to higher surface densities and mass accretion rates that lead to a hotter mid plane and a more distant snow line. Higher mass accretion rates and later times have the opposite effect, leading to a colder, optically thinner disk with a closer-in snow line. In this regime, irradiation may become important as a heating source in the mid plane.

\section{Implications for water delivery}\label{sec:nbody}
To explore how the range in snow line locations may impact the water content of terrestrial planets in the habitable zone, we use the set of N-body simulations from \cite{Ciesla:2015ha}. This approach follows that of \cite{2006Icar..184...39O}, \cite{2007ApJ...669..606R} and \cite{2007ApJ...660L.149L}, for forming terrestrial planets via giant impacts as in the solar system. The simulations are started from a $5 M_\oplus$ disk with equal mass planetary embryos (1/20 Earth mass) and planetesimals (1/20 embryo mass), spaced such that they define a $1/R$ power law in surface density, as in \cite{2006Icar..184...39O}, between 0.5 and 4.0 au. We do not include giant planets, but note that  they may be rare around low-mass stars \citep{Johnson:2010gu} and water delivery can take place in their absence \citep{Quintana:2014hr}. Four simulations are run with near-identical initial conditions to take into account stochastics. These are repeated for lower stellar masses (0.2, 0.4, 0.6 and 0.8 $M_\odot$) by scaling all masses (total, embryo, and planetesimal) to the stellar mass and adjusting the inner and outer edge to encompass both snow line and habitable zone. After 200 Myr, a number of terrestrial planets has formed in the inner regions, see \cite{Ciesla:2015ha} for details. 

\begin{figure}
	\includegraphics[width=\linewidth]{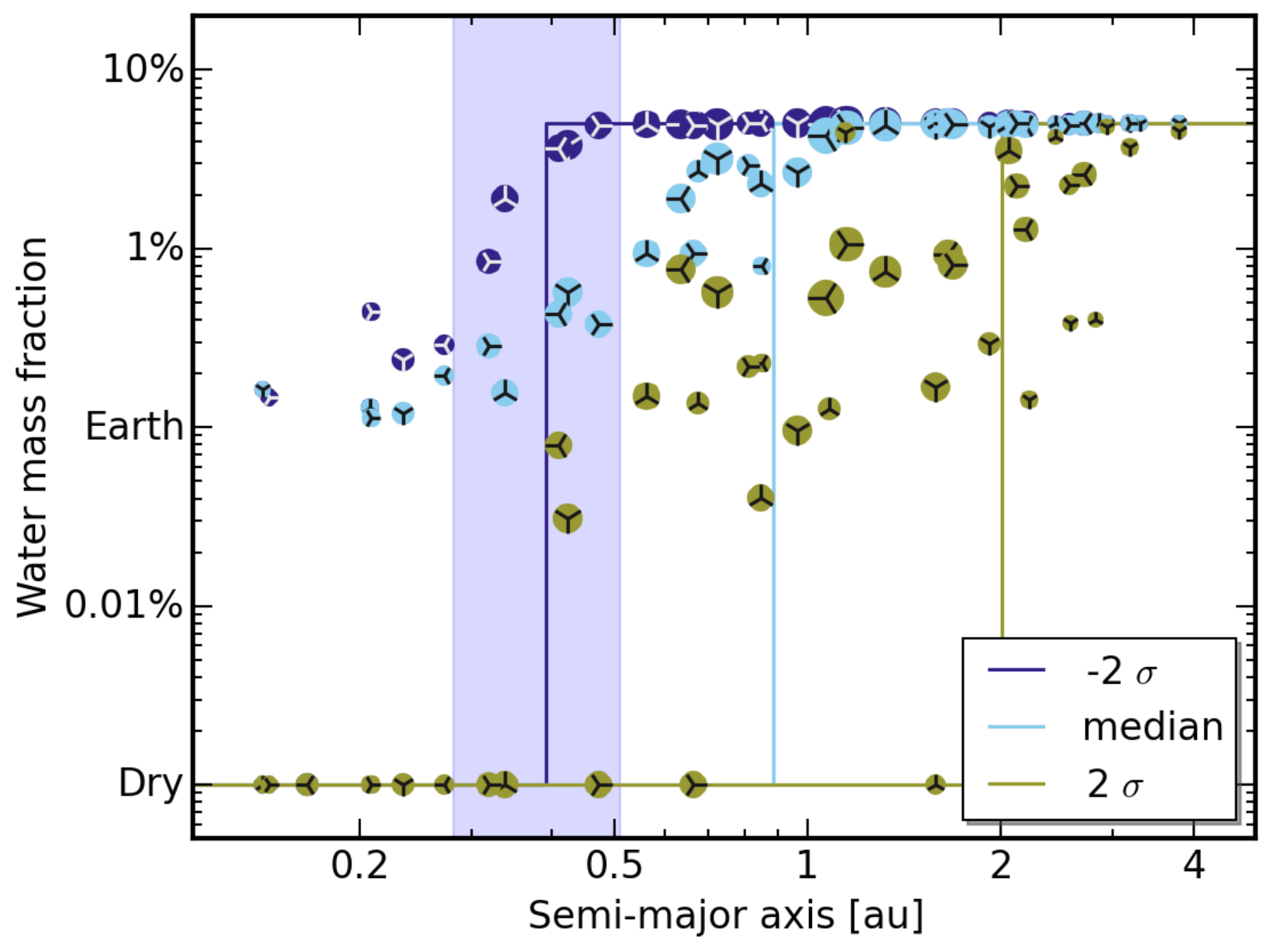}
	\caption{Planetary water abundances for a $0.6 M_\odot$ star, computed for three different snow line locations (median and $\pm 2 \sigma$). Compare with CC case in figure 6 in \cite{Ciesla:2015ha}. 
	\label{fig:ice_SL}
	}
\end{figure}

The water content of these planets depends on the composition of accreted planetary embryos and planetesimals, which in turn depends on their starting locations. We assume that the division between icy and rocky building blocks is located at the snow line, i.e. that planetesimals and embryos do not move radially during their growth process, and that the transition from pebbles to planetesimals and embryos is short compared to the time scale at which the snow line moves inward \citep[e.g.][]{Carrera:2015vy}. Assuming the water content of icy building blocks is low (5\%, similar to carbonaceous chondrites), the extra mass outside of the snow line does not affect the dynamics of the system, allowing the snow line to be inserted \textit{a posteriori}. By keeping track of the starting locations of planetary building blocks that comprise the final planets, their water content can be calculated for different snow line locations from a single simulation. Figure \ref{fig:ice_SL} gives an example of this approach, which shows the water content of planets in the $0.6 ~M_\odot$ simulations calculated from three different locations of the snow line. 

\begin{figure*}
	\includegraphics[width=\figwidth\linewidth]{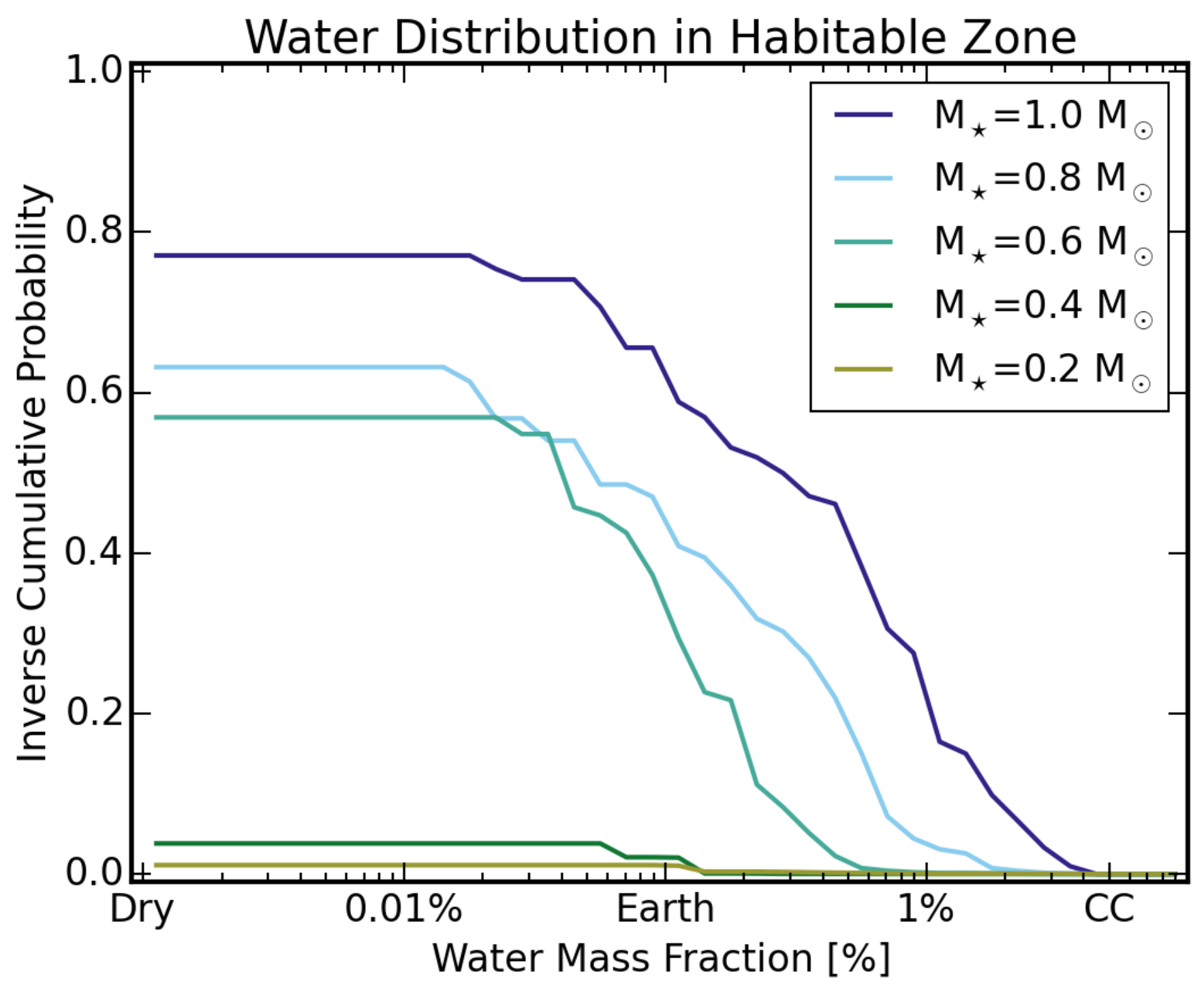}
	\includegraphics[width=\figwidth\linewidth]{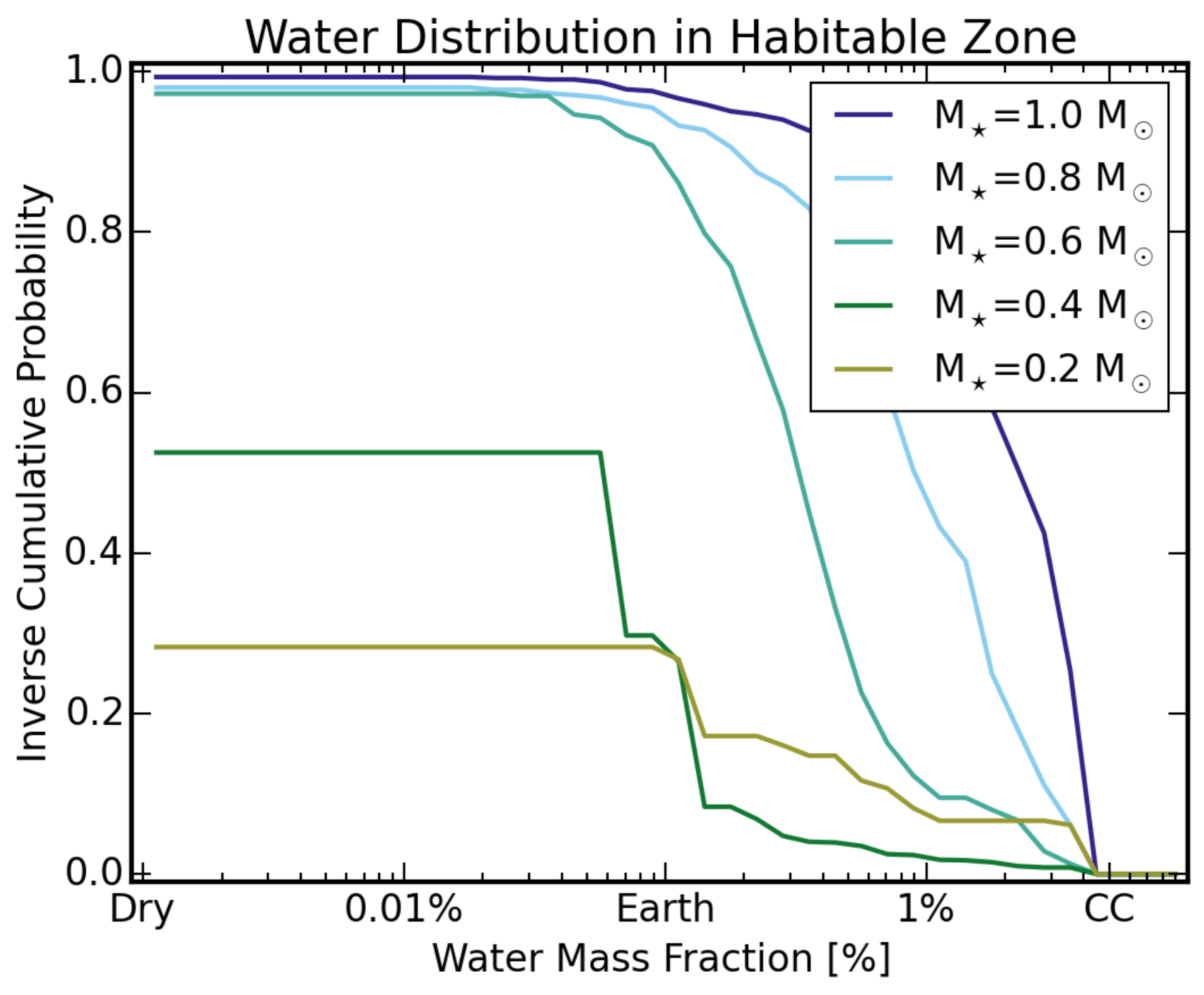}

	\caption{Fraction of planets in the habitable zone with at least a given mass fraction of water, for stars of different masses. The \figa and \figb panel use a snow line calculated with ISM-like and larger grains, respectively. 
	\label{fig:PDF_HZ}
	}
\end{figure*}

By calculating the water content of planets in the habitable zone for a range of snow line locations from the simulations, and assigning to each snow line a probability based on the observed dispersion in mass accretion rates (Eq. \ref{eq:lognormal}, see also Fig. \ref{fig:diag}), we calculate the probability distribution of the water fraction of planets in the habitable zone. These are shown as \textit{cumulative} probability distributions in Fig. \ref{fig:PDF_HZ}.

We take a water mass fraction of 0.1\% as a reference point for early Earth, between 0.03-0.1\% currently in the mantle (\citealt{lecuyer2013water}) and as much as 1\% in the past according to \cite{Abe:2000vr}.
In the simulations with sun-like stars, for a snow line set by small grains, 20\% to 40\% of planets in the habitable zone remain dry, as a fraction of stars has a mass accretion rate high enough to put the snow line too far away from the habitable zone ($~4$-$5$ au) to allow water delivery via giant impacts. The fraction of dry planets is higher in simulations with lower-mass stars, reaching 40\% to 60\% for 0.6 and 0.8 solar mass stars, and close to 100\% for $<0.4$ solar mass stars, in line with previous results (\citealt{2007ApJ...669..606R,Ciesla:2015ha}). The one-sigma range of water fractions spans more than an order of magnitude, significantly larger than the stochastic dispersion in the N-body simulations (a factor of $\sim$2 to $\sim$5).

Using larger grains for calculating the snow line location significantly changes the results. Because the snow line is located closer to the star, water delivery is more efficient: the fraction of dry habitable-zone planets in our simulations decreases to zero for 0.8 and 1.0 solar mass stars, and to 10\% for 0.6 $M_\odot$. A small fraction of habitable zone planets around low-mass stars does receive earthlike amounts of water, around 20\% for 0.2 $M_\odot$ and 0.4 $M_\odot$.

\section{Discussion and Conclusions}\label{sec:discussion}
We derived the fraction of water-bearing terrestrial planets in the habitable zone by assuming they are formed in-situ. There is, however, growing observational evidence that migration of planets or their building blocks plays an important role in the formation of super-Earths and mini-Neptunes at short orbital periods \citep{2012ApJ...751..158H,2013ApJ...764..105S, 2014MNRAS.440L..11R,Mulders:2015ja}. Whether this mechanism  is able to form smaller, earth-sized planets farther out remains an open question. If this were the case, the population of terrestrial planets in the habitable zone might be a mixture of planets formed in-situ and through migration. As migrating planets typically form beyond the snow line \citep[e.g.][]{2014A&A...569A..56C,Izidoro:2014eg}, they are expected to have water mass fractions of order 50\% \citep{Ogihara:2009cu, Tian:2015bx}. Hence, the fractions of wet terrestrial planets quoted in this paper can be considered as lower limits.

An additional concern for the habitability of habitable zone planet around lower mass stars is that water loss may may be more efficient as these planets form hotter than those around sunlike stars \citep{2007ApJ...660L.149L}. \cite{Ramirez:2014bx} show that duing the bright pre-main sequence phase of low-mass stars, stellar fluxes are high enough to trigger a runaway green house effect that leads to enhanced water loss. \cite{Tian:2015bx} come to a similar conclusion by modeling water los through hydrodynamc escape in a planet population synthesis model. 
For a planet to remain habitable, water has to arrive later during the pre-main-sequence evolution of the star.
The timing and delivery mechanism of water may be crucial here: In our simulations, water delivery persists over $\sim 100$ Myr time scales. In addition, water is not delivered in the form of ices that would be directly deposited in the atmosphere. Rather, it arrives as hydrated minerals that would enter the atmosphere later via vulcanic outgassing. A detailed study of arrival times and atmospheric release of water may be neccessary to address whether water can be deliverd late enough and in sufficient quantities to avoid escape.

Despite water delivery to low-mass planets in the habitable zone being less efficient than around sunlike stars \citep{2007ApJ...660L.149L,2007ApJ...669..606R}, we show that a small fraction may still receive earthlike amounts of water due to the dispersion in snow line locations. When taking into account the larger number of M stars with respect to G stars (factor ten), their higher planet-occurrence rates (factor two, \citealt{Mulders:2015ja}), and increased water delivery from more comet-like icy bodies \citep{Ciesla:2015ha}, the majority of water-rich terrestrial planets may still be found around low-mass M stars.

\ifemulate
	\ifastroph
		\bibliography{snowline.bbl}
	\else	
		\bibliography{papers3,books}
	\fi
\else	

\fi


\appendix

\section{The role of irradiation in determining the snow line location.}\label{app:mcmax}

\begin{figure}
	\includegraphics[width=\linewidth]{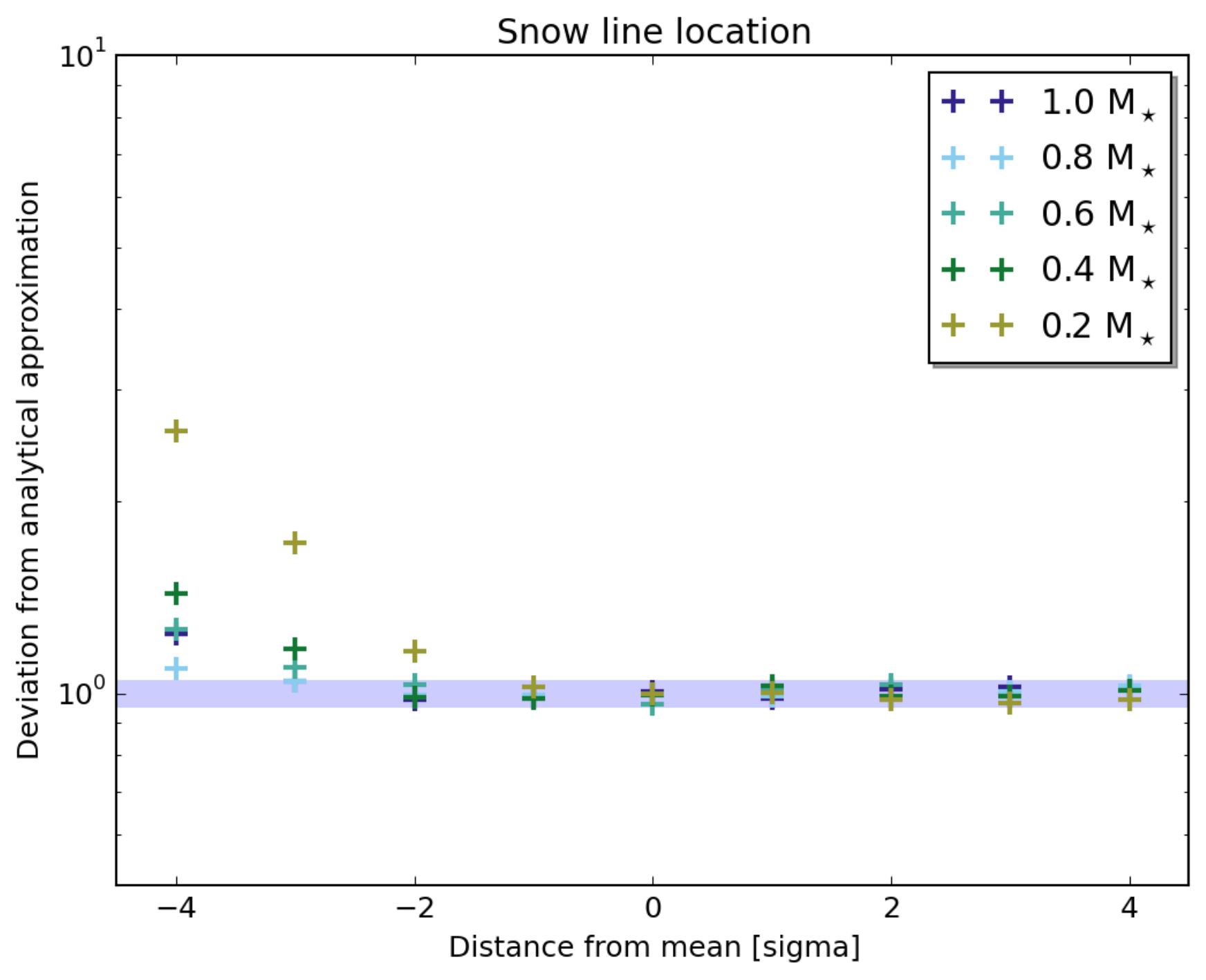}
	\caption{Ratio of the snow line location calculated from radiative transfer models with respect to the analytical approximation (Eq. \ref{eq:analytical}). The blue shaded region indicates a $10\%$ error. 
	The analytical approximation shows good agreement with the radiative transfer models over the $\pm4 ~\sigma$ range of mass accretion rates for stellar masses $\gtrsim 0.6 M_\odot$. At lower stellar masses, irradiation leads to a larger snow line at mass accretion rates $3$ and $2$ sigma below the median mass accretion rate for a stellar mass of $0.4 M_\odot$ and $0.2 M_\odot$, respectively.
	\label{fig:SL_MCMax}
	}
\end{figure}

Throughout this paper, we have assumed that the midplane temperature at the location of the snow line is determined by viscous heating (Eq. \ref{eq:analytical}), and irradiation is negligable. For sunlike stars, \cite{2011Icar..212..416M} have shown that this is a good approximation for the mass accretion rates considered in this paper (See also \citealt{2007ApJ...654..606G,Oka:2011jh}), assuming accretion is not layered \citep[e.g.][]{2011ApJ...740..118L}. Around lower mass stars, lower mass accretion rates and a brighter pre-main sequence phase may lead to a larger role of irradiation. To verify the assumption that equation \ref{eq:analytical} also holds for low-mass stars, we compute the original radiative transfer model from \cite{2011Icar..212..416M} for lower-mass stars. For the stellar photosphere, we use the \cite{1998A&A...337..403B} evolutionary tracks at $t=1$ Myr. We explore a $\pm 4\sigma$ range in mass accretion rates for the stellar masses used in this paper. Figure \ref{fig:SL_MCMax} shows the location of the snow line (i.e., the location where half the water is condensed into ice) divided by the predicted location from Eq. \ref{eq:analytical}. 

For stellar masses larger than $0.4 M_\odot$, the location of the snow line is predicted accurately down to two sigma below the median within the precision of the radiative transfer model ($\sim 5\%$). At smaller mass accretion rates, irradiation becomes imporant and equation \ref{eq:analytical} starts underpredicting the location of the snow line. For the $0.2 M_\odot$ case, irradiation has a larger influence, and the predicted location of the snow line starts deviating at two sigma below the median.

\end{document}